# Computer Navigated Spinal Surgery Using Magnetic Resonance Imaging and Augmented Reality

Songyuan Lu*'[1], Jingwen Hui*'[2], Jake Weeks[2], David B. Berry[3], Fanny Chapelin[2,4], Frank Talke[1,+]

University of California, San Diego
[1]Department of Mechanical and Aerospace Engineering
[2]Shu Chien - Gene Lay Department of Bioengineering
[3]Department of Orthopaedic Surgery
[4]Department of Radiology

**Abstract:**

Current spinal pain management procedures, such as radiofrequency ablation (RFA) and epidural steroid injection (ESI), rely on fluoroscopy for needle placement which exposes patients and physicians to ionizing radiation. In this paper, we investigate a radiation-free surgical navigation system for spinal pain management procedures that combines magnetic resonance imaging (MRI) with fiducial ArUco marker-based augmented reality (AR). High-resolution MRI scans of a lumbar spinal phantom were obtained and assembled as a surface mesh. Laplacian smoothing algorithms were then applied to smoothen the surface and improve the model fidelity. A commercially available stereo camera (ZED2) was used to track single or dual fiducial ArUco markers on the patient to determine the patient's real-time pose. Custom AR software was applied to overlay the MRI image onto the patient, allowing the physician to see not only the outer surface of the patient but also the complete anatomy of the patient below the surface. Needle-insertion trials on a 3D-printed 3-vertebra phantom showed that dual-ArUco marker tracking increased the accuracy of needle insertions and reduced the average needle misplacement distance compared to single-ArUco marker procedures. The average needle misplacement is comparable to the average deviation of 2 mm for conventional epidural techniques using fluoroscopy. Our radiation-free system demonstrates promise to serve as an alternative to fluoroscopy by improving image-guided spinal navigation.

**Introduction:**

Low back pain (LBP) is one of the leading causes of reduced productivity, particularly affecting the working population due to poor posture and prolonged inactive behaviors [1]. Spinal pain management procedures, such as radiofrequency ablation (RFA) [2] and epidural steroid injections (ESI) [3], have become increasingly more prevalent to alleviate pain and restore mobility impaired by neural compression or spinal inflammation. Precise needle placement is critical in these minimally invasive procedures. However, even in controlled laboratory settings, needle-insertion misplacements can range from less than 1 mm to nearly 6 mm per 50 mm of traversed tissue, depending on the operator's skill and needle type [4]. Kopacz showed that Tuohy needles, used specifically for spinal pain management procedures, deflected 1.4-2.1 mm when using a 22-gauge needle and 2.6-3.6 mm when using an 18-gauge needle during a 5 cm penetration of body tissue [4]. These procedures are typically performed under fluoroscopic guidance, which provides real-time imaging but exposes both patients and clinicians to ionizing radiation. This exposure carries long-term health risks, including increased likelihood of cancer, radiation-induced tissue damage, and potential genetic harm [5].

Recent investigations have explored alternative navigation techniques. For instance, optical tracking technologies, employing stereo cameras (ZED2, StereoLabs) and reflective markers, have demonstrated the potential to eliminate radiation exposure while maintaining guidance accuracy [6]. Also, robotic platforms with integrated visual feedback have been developed to assist spinal surgeries [7]. In addition, radiation-free procedures with MRI guidance have been investigated using specialized instruments [8]. Furthermore, augmented reality (AR) has emerged as a promising radiation-free surgical guidance technology that is capable of overlaying virtual anatomical structures directly onto the patient's body during procedures [9]. For instance, AR systems have been used to improve catheter placement in MRI-guided cardiac ablation therapy without the use of fluoroscopy [10]. Similarly, an AR-based neurosurgery platform employing optical tracking has demonstrated highly accurate alignment of virtual anatomy in endoscopic procedures [11]. Examples of commercially-available AR navigation systems for spinal procedures include Augmedics Xvision (Augmedics, Inc.) and the NuVasive Pulse Platform (NuVasive, Inc.). Both systems depend on fluoroscopy for real-time anatomy visualization and on infrared markers for registration. Despite these advances, current AR implementations still encounter critical limitations, including insufficient patient-specific anatomical accuracy, and reliability in maintaining real-time alignment during dynamic surgical conditions [12].

To address these limitations, Hui et al. investigated an augmented reality (AR) navigation system which leverages magnetic resonance imaging, single-ArUco marker tracking, and computer vision techniques for minimally invasive spinal procedures [13]. In Hui et al.'s approach, a 3D computer aided design (CAD) spinal model was overlaid on the patient using AR and a single fiducial ArUco marker [13]. The use of a spinal CAD model was a shortcut that allowed testing of the general approach without the clinical complications arising from MRI acquisitions. Preliminary results showed great promise of the proposed approach.

In this study, we present an improved, MRI-based computer navigation system, starting with the magnetic resonance image acquisition to deliver practical and clinical-grade guidance for spinal procedures. The new system proposed here includes the following four key stages, as illustrated in Figure 1.
**Stage 1 (MRI Acquisition with Fiducials):** Disposable MRI‑visible fiducial markers are placed by the surgeon on the patient, in our case a spinal phantom, near the target anatomical region (Fig. 1a). The phantom then undergoes an MRI scan with the MRI fiducial markers and the ArUco markers in place.
**Stage 2 (3D Model Reconstruction and Enhancement):** Using the MRI scans, we construct a high-resolution, 3D spinal model of the phantom through image segmentation (Fig. 1b). To reduce discretization-induced step changes on the model, we employ a Laplacian smoothing algorithm that iteratively refines the mesh, reduces artifacts, and enhances the anatomical accuracy and surface smoothness.
**Stage 3 (AR Overlay Preparation and Registration):** Prior to the surgical procedure, the AR fiducial (ArUco) markers must be placed at the original MRI fiducial locations (Fig. 1c). This preserves the marker-spine spatial relationships established during MRI scanning. A stereo camera (ZED2, StereoLabs) detects the ArUco markers in real-time and derives the orientation and position of the patient's spine based on reference points and positional information from the MRI fiducials. This process facilitates dynamic AR overlay of the reconstructed 3D spine onto the patient's back (in our case, the phantom) via an AR headset (Meta Quest 3, Meta).
**Stage 4 (Surgical Application and Validation):** The surgeon performs the needle insertion guided by real-time AR visualization, eliminating ionizing radiation (Fig. 1d). We validate the system through controlled needle-insertion experiments on our spinal phantom model.

The integration of MRI-based modeling of a patient or phantom, the applications of advanced smoothing algorithms, and the improved dual-ArUco marker AR tracking system implemented in this investigation represent important advancements towards a clinically functional model. Details of this approach will be discussed in the following sections.

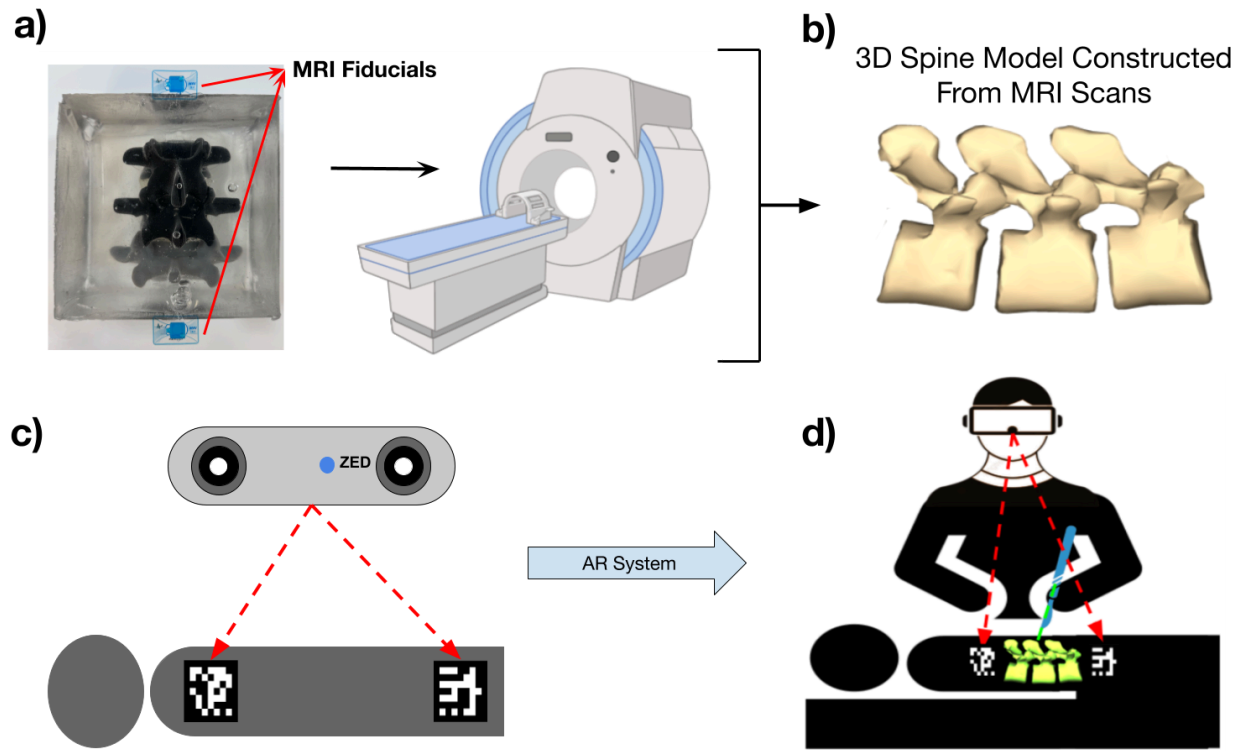

**Fig. 1:** Overview of computer navigation system. a) MRI scans of patient or phantom b) 3D spine model obtained from MRI scans for overlay on patient or phantom c) Stereo camera to track the real-time positions of the ArUco markers, and d) Projection of AR spine model from MRI scans onto the patient via AR headset.

**Experimental Methods and Procedure:**

Phantom Model Fabrication:
A 3-vertebra spinal phantom was 3D-printed from an open-source CAD model and embedded in tissue-mimicking ballistic gel (Fig. 2) [14]. MRI-visible fiducial markers (LiquiMark 10 mm Square Liqui-Pouch, Suremark) were placed on the surface of the ballistic gel marking the area of interest. The MRI visible markers are needed to provide spatial information required for the AR overlay. The 3D phantom was used to perform controlled needle-injection experiments for validation.

MRI Scanning Procedure of the phantom model:
In Fig. 2, the spinal phantom used in this work is shown. The phantom consists of three vertebrae of the lumbar spine, 3D printed from an open-source CAD file. A high‑resolution T1‑weighted Dixon MRI sequence with an image resolution of 1mm x 1mm x 1mm was acquired on a 3T MRI scanner (Prisma, Siemens Healthineers, Erlangen). The scans were acquired with in-phase and out of phase echos to give robust water-fat separation. Pulse sequence parameters can be found in supplementary table 1. We performed slice-by-slice segmentation of the raw DICOM volumes using a commercially available software (ITK-SNAP) to isolate the vertebral bodies, intervertebral discs, and surrounding bony landmarks. The segmented volumes were then converted into a single surface mesh and individual components (vertebrae, discs, etc.) were fused into a single contiguous stereolithography (STL) model.

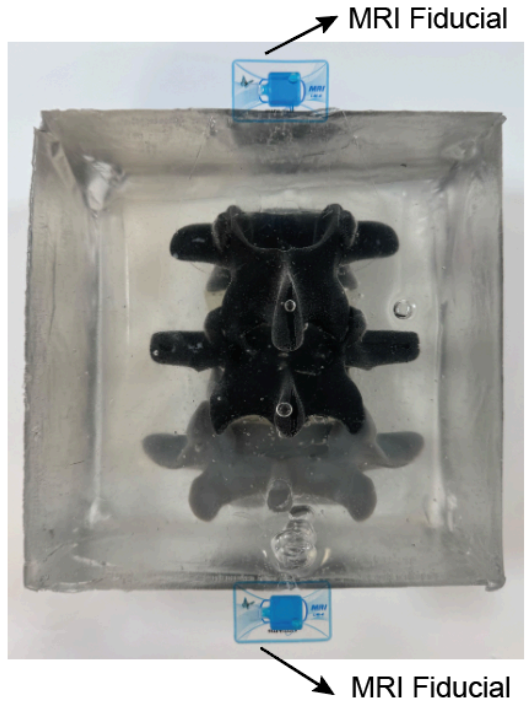

**Fig. 2:** Top view of the spinal phantom with MRI fiducials attached.

Augmented Reality System:

We use AR for our computer navigation system because it enables real-time high-resolution visualization of the internal anatomy without radiation [15]. Our AR system captures spatial information of the patient (in our case the vertebral phantom) via fiducial ArUco marker registration and overlays the patient-specific virtual spine model directly onto the patient's back (or the phantom model's back). Among the various markers, we chose ArUco markers because they are non-invasive and consistently perform at submillimeter accuracy [16]. We evaluate our advanced AR system through a series of surgical injection simulation experiments.

System Architecture:
The AR system consists of two main components:
1. ArUco marker detection using a pre-calibrated stereo camera (ZED2, StereoLabs) and spatial pose computation using real-time RGB and depth data processed via Vision Software Development Kit (ZED SDK and AI Vision Framework).
2. AR visualization delivered through a 3D application (Unity) on an AR headset (Meta Quest 3, Meta).

A custom tracking algorithm (Fig.3, algorithm1) was implemented within Unity that allows real-time communication with the stereo camera (ZED2) and seamless integration of the patient-specific spine model.

**Algorithm 1** Real-Time Marker-Based AR Registration and Tracking

**Input:** Stereoscopic image stream from a ZED camera; predefined ArUco marker dictionary; camera intrinsic and extrinsic parameters
**Output:** Continuously updated pose-aligned AR content overlaid based on real-world marker positions

1: **Initialize:**
2:     Configure and start the ZED stereo camera.
3:     Load camera calibration parameters (intrinsics and extrinsics).
4:     Import the ArUco marker dictionary and define physical marker dimensions.
5:     Register virtual objects with corresponding marker IDs using `MarkerObject` instances.
6: **while** the image stream is active **do**
7:     **Acquire Frame:** Capture a stereo frame and convert it to OpenCV-compatible format.
8:     **Detect Markers:** Apply ArUco detection to identify all visible markers.
9:     **for** each detected marker **do**
10:         Estimate the 6-DoF pose of the marker using corner coordinates and camera intrinsics.
11:         Retrieve the corresponding `MarkerObject` based on the marker ID.
12:         **if** a matching object is found **then**
13:             Update the object's position and orientation with pose smoothing (e.g., low-pass filtering).
14:             Ensure the object is marked as active and is rendered in the AR scene.
15:         **end if**
16:     **end for**
17:     **Handle Occlusion:**
18:     **for** each registered `MarkerObject` not detected for $T_{\text{miss}}$ consecutive frames **do**
19:         **if** auto-disable is enabled **then**
20:             Deactivate the object to prevent visual clutter from stale overlays.
21:         **end if**
22:     **end for**
23: **end while**

**Fig. 3:** Marker Tracking algorithm

Marker Tracking and Pose Computation:
The ArUco markers were placed on the patient's back or on the phantom at the locations of the MRI fiducials. The stereo camera detects these markers and computes a six degree of freedom (6-DOF) pose estimation for each marker. Pose data is used in real time to update the position, orientation, and scale of the 3D spinal model reconstructed from MRI data. A Unity-based AR application is used to handle the data integration and alignment of the virtual spinal model onto the patient's anatomy. This virtual spine is dynamically scaled based on the distance between the ArUco markers and the stereo camera as the patient or the camera moves.

Tracking Modes:
In this paper, we compare the dual-ArUco marker tracking algorithm with the single-ArUco marker tracking algorithm used in [13]. Both methods are assessed for tracking accuracy in needle-insertion experiments.

<u>Single-ArUco Marker Tracking</u>: In this setup, the algorithm continuously receives spatial information from the ZED2 camera and simultaneously processes the location data and outputs the calibrated position and orientation of one marker. Based on the marker position, the system then modulates and updates the virtual 3D spine model's location in real time. Additionally, the scale of the virtual spine is adjusted based on the distance between the marker and the camera, ensuring that the displayed spine changes size appropriately.

<u>Dual-ArUco Marker Tracking</u>: In the dual-marker detection version, the system detects the relative positions of both ArUco markers, thus deriving the computed spine coordinates and rotation angles based on the spatial relationships from MRI data. The distance between the center of the markers is used to dynamically scale the virtual 3D spine, thereby adjusting its size as the markers move closer to or farther from the camera. If one marker is temporarily blocked from the ZED camera vision (e.g. obstructed by the surgeon), the dual-marker tracking mode automatically converts back to single-marker tracking mode. The system automatically reverts back to dual-marker tracking as soon as the camera regains full vision of both markers.

AR Output Visualization:
The AR visualization is overlaying the virtual 3D spine model onto the physical model (Fig.4). This view is displayed to the surgeon via a headset (Meta Quest 3, Meta) in real time. The real time visualization of the patient's spine anatomy enhances clinical accuracy by constantly tracking and displaying the surgical area of interest. Fig.4.a) shows the overlay of the AR spine onto the physical model. Fig.4.b) shows the view of the surgeon during a procedure with the surface covered by silicone-based material which imitates human skin.

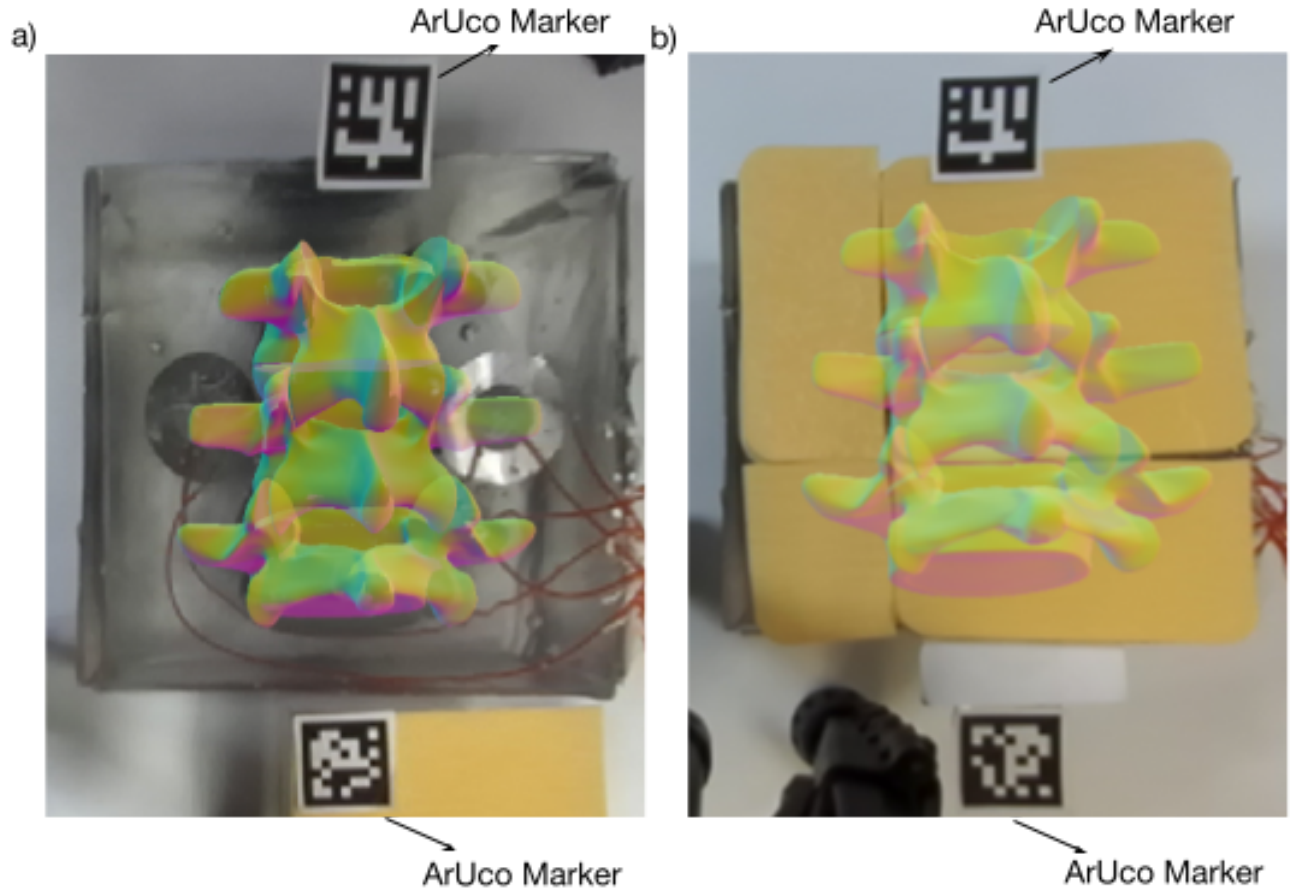

**Fig. 4:** AR spinal overlay. a) Phantom of spine with AR overlay b) Surgeon's actual view during procedure (the real spine is covered up by an opaque surface sheet simulating skin).

**Experimental setup:**

MRI:

The MRI segmentation process involves stitching together many planar cross‑sections, producing a surface mesh that exhibits "stair‑step" artifacts and choppy edges at boundaries. Because spinal pain management procedures require accurate visualization of the spine's anatomy, we need to apply a smoothing procedure and evaluate how close the CAD profile of the MRI generated 3D model is to the actual phantom. To smoothen the rough exterior produced by the assembly of MRI cross sections, we applied an explicit Laplacian smoothing algorithm as described in the next section.

MRI Smoothing Optimization (Laplace Smoothing):
The Laplacian smoothing operation uses a mesh-smoothing filter that moves each vertex toward the average of its neighbors, thereby reducing surface noise while retaining the overall shape of the model [17]. A pseudo code for the Laplacian smoothing algorithm used in this paper is shown in algorithm 2 (Fig. 5). At every iteration, each vertex of the mesh moves towards the average position of its k nearest neighbors, weighted by a smoothing factor α. This operation can be described by the following equation:

$$(1): v_i^{(t+1)} = v_i^{(t)} + \alpha\left(\frac{1}{|N_k(i)|} \sum_{j \in N_k(i)} v_j^{(t)} - v_i^{(t)}\right)$$

where $v_i^{(t)}$ is the 3D position of vertex i before the t-th update, and $v_i^{(t+1)}$ is the new position of the vertex i at the t-th iteration, $\alpha$ is a smoothing factor that controls how far each point moves towards its neighbor, $N_k(i)$ is the set of all k neighbors of i, k is the neighbor count, and t is the number of iterations.

To quantify the smoothing problem, it is common to introduce the so-called Dice index, given by [18]:

$$(2): Dice_{Shell} = \frac{2|A_{shell} \cap B_{shell}|}{|A_{shell}| + |B_{shell}|}$$

where $A_{shell}$ and $B_{shell}$ represent single-voxel-thick surface shells that were extracted from the 3D mesh.

---

**Algorithm 2** Laplacian Smoothing Optimization

---

**Input:** Phantom Mesh (Ground Truth Mesh (GT)), MRI Mesh, Resolution $R$, Neighbor Counts $K$, Iteration Counts $I$, Weights $\alpha$
**Output:** Top $N$ smoothed mesh by Dice score

1: Compute shared bounding box of GT and MRI mesh
2: Voxelize GT mesh at resolution $R$ to obtain $GT\_surface$
3: Initialize empty list RESULTS
4: **for** each $k$ in $K$ **do**
5:     **for** each $it$ in $I$ **do**
6:         **for** each $\alpha$ in $A$ **do**
7:             Apply Laplacian smoothing to MRI mesh with $(k, it, \alpha) \rightarrow SM\_mesh$
8:             Voxelize $SM\_mesh$ at resolution $R$ using shared bounds $\rightarrow SM\_vol$
9:             Compute Dice coefficient $d = Dice(GT\_surface, SM\_vol)$
10:            Add tuple $(d, k, it, \alpha, SM\_mesh)$ to RESULTS
11:         **end for**
12:     **end for**
13: **end for**
14: Sort RESULTS by $d$ in descending order and take first $N$ entries $\rightarrow$ BEST
15: **for** each entry in BEST, with rank $r$ **do**
16:     Export mesh as "rank⟨r⟩_k⟨k⟩_iters⟨it⟩_α⟨α⟩_dice⟨d⟩.stl"
17:     Print rank $r$, Dice score $d$, and settings $(k, it, \alpha)$
18: **end for**

---

**Fig. 5:** Laplacian smoothing optimization algorithm

Evaluation:

We grid-searched neighbor counts k∈{8,16,32,64,128}, iteration counts i∈{1,5,10,20,50}, and weights α∈{0.1,0.3,0.5,0.7,1.0}. For each (k,i,α) triplet, we computed a 3D surface-shell Dice index between the smoothed MRI mesh and the CAD phantom mesh and picked the parameters that maximized the Dice index.

Augmented Reality:

Using the spinal phantom model, we were able to collect accurate insertion experiment performance data of our AR system without having to test on actual patients. This reduced the time and costs without compromising the accuracy of the procedure.

Physical Spine Model Design:
The ArUco marker placement for both single-ArUco marker tracking and dual-ArUco marker tracking setups is shown in Fig.6a and Fig. 6b, respectively. During the needle insertion experiments, the top surface of the spinal phantom was covered with a skin-like material to simulate a real patient in a real procedure. The target of the surgeon is the location of the medial branch nerve on the second vertebra which is a common target point of radiofrequency ablation (RFA) and epidural steroid injection (ESI) procedures [19]. To detect how close the placement of the needle was to the target area, we used a specially designed circuit as shown in Fig.7.

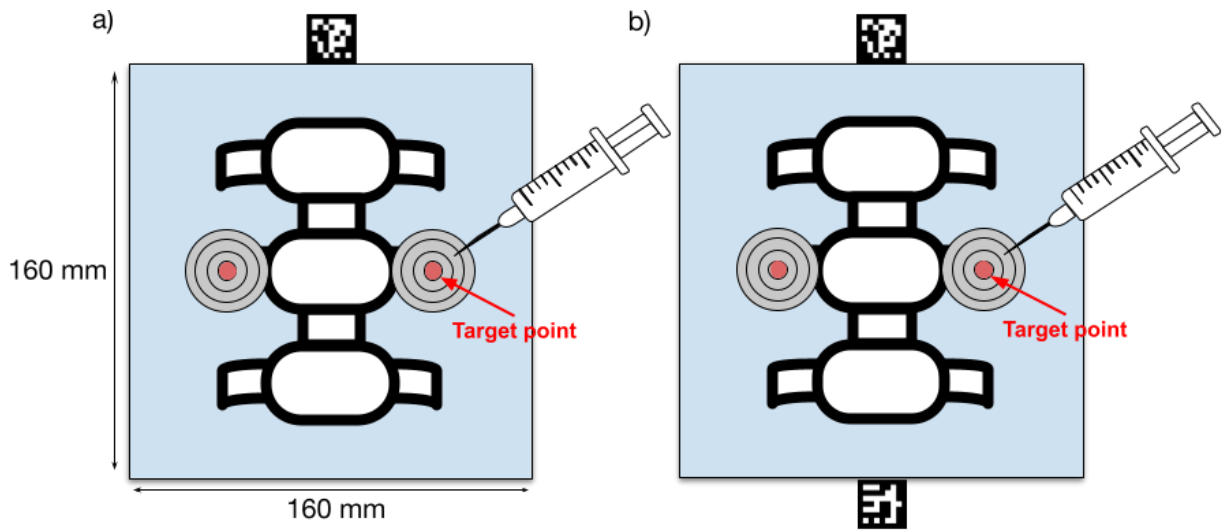

**Fig. 6**: Top view of phantom setup. a) Single-ArUco marker setup and b) Dual-ArUco marker setup.

Circuit Design:
Fig. 7a shows the schematic of the circuit design used to detect the accuracy of needle placements during simulated medical procedures. Fig. 7b shows the detector, consisting of stacked aluminum rings with LEDs that light up on contact to display the position of needle placement on any of the four rings. The radius of the smallest ring is 1 mm. This value is smaller than the misplacement error of 2mm that is typically seen in conventional spinal procedures. We chose the inner radius to be 1mm to enhance the targeting accuracy of our measurement.

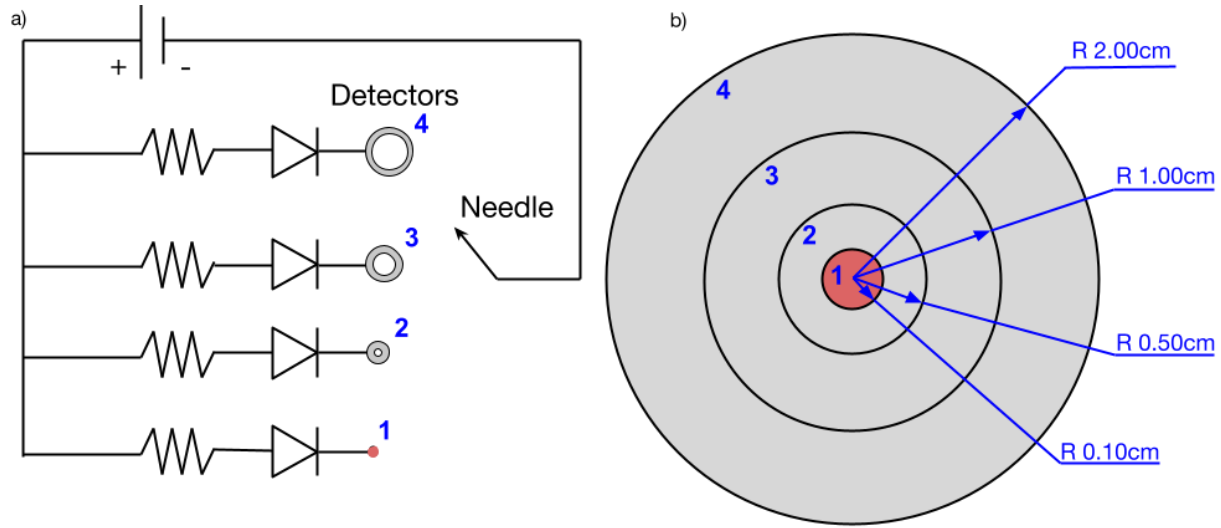

**Fig. 7**: Feedback circuit design. a) Circuit diagram of the detector circuit b) Dimensions of detectors consisting of four aluminum rings.

Test procedure:
50 simulated surgical insertion trials were performed for both the single-ArUco marker setup and the dual-ArUco marker setup, using a) the CAD virtual overlay model, b) the unsmoothed MRI-virtual overlay model, and c) the MRI-smoothed virtual model, overlaid on the physical phantom. Another 50 trials were performed serving as baseline without any ArUco markers or AR guidance. We then evaluated the results for each experiment by analyzing the deviation of the insertions from the target.

**Results and Discussion:**

MRI Results:

Our investigation showed that the unsmoothed MRI model had a surface-shell Dice coefficient of 78.4%, while the smoothed MRI model showed a surface-shell Dice coefficient of 86.6%. Fig. 8a shows the phantom CAD spinal model, Fig.8b shows the unsmoothed MRI model, and Fig. 8c shows the MRI model after Laplacian smoothing. We observe clearly the discretization error in Fig. 8b, and the improvements in Fig. 8c brought on by the smoothing process.

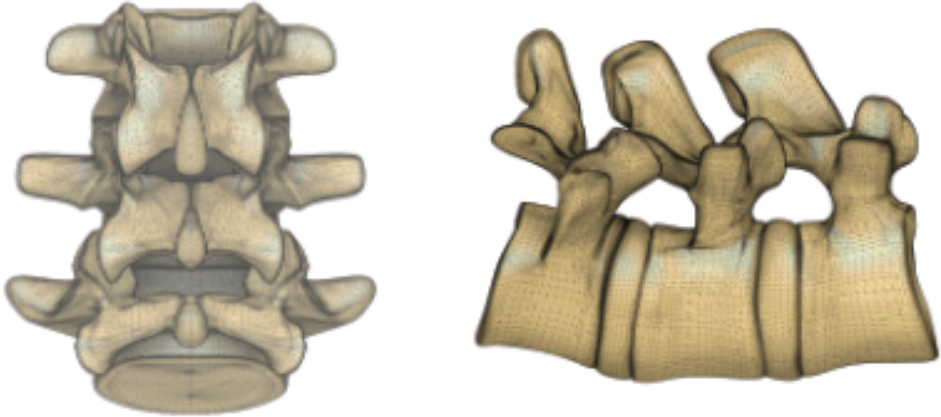

a) Spinal CAD model

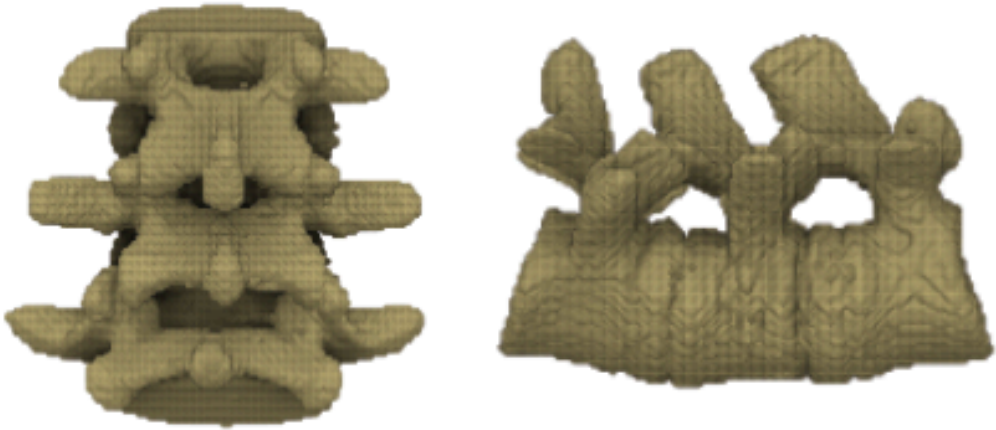

b) MRI Model From Scans

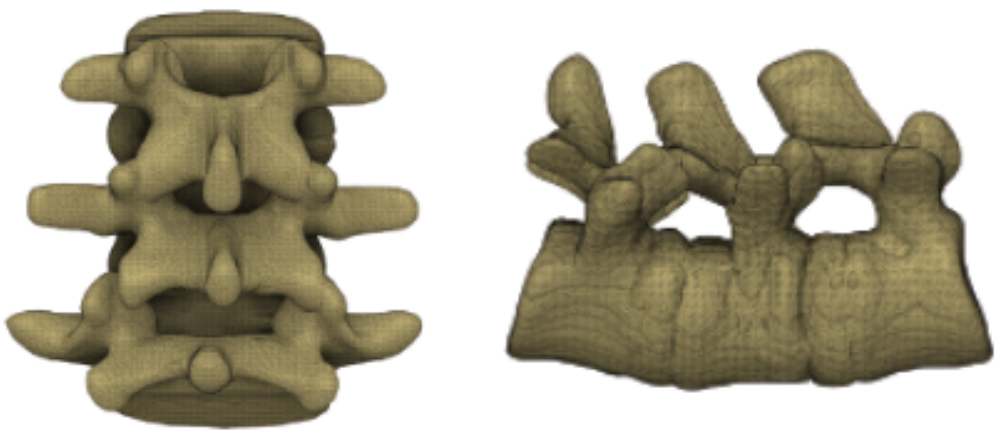

c) Smoothed MRI Model

**Fig. 8:** Spine Models. a) CAD spinal model. b) MRI-scanned spinal model. c) smoothed MRI model.

Augmented Reality (AR) Results:

To evaluate the accuracy of the needle placement, the radius of each detector ring was measured so that an activated ring indicates how far the needle was misplaced from the target. We define a "high-accuracy insertion" as one that falls within the innermost detector ring (radius ≤ 1 mm). Insertions in the outer detector rings were assigned misplacement values equal to their radial distance from the target point.

Fig. 9 shows the comparative performance of the single-ArUco marker and dual-ArUco marker AR tracking system for the CAD phantom model, the raw MRI model and the smoothed MRI model. For all three models, the dual-marker system was found to increase the "high-accuracy insertion rate" and reduce the average needle misplacement relative to the single-marker system. Specifically, the smoothed MRI model for the dual-ArUco marker setup showed a "high-accuracy insertion" rate of nearly 60% with an average deviation of just 1.9 mm. In comparison, the single-marker system had a "high-accuracy" insertion rate of 46% with an average misplacement of approximately 2.5 mm. This demonstrates clearly the enhanced accuracy of dual-marker tracking, since the system consistently achieved a performance comparable to the ~2 mm deviation associated with conventional fluoroscopy methods [4]. Our dual-marker AR system thus represents a robust and radiation-free alternative to traditional fluoroscopic navigation systems.

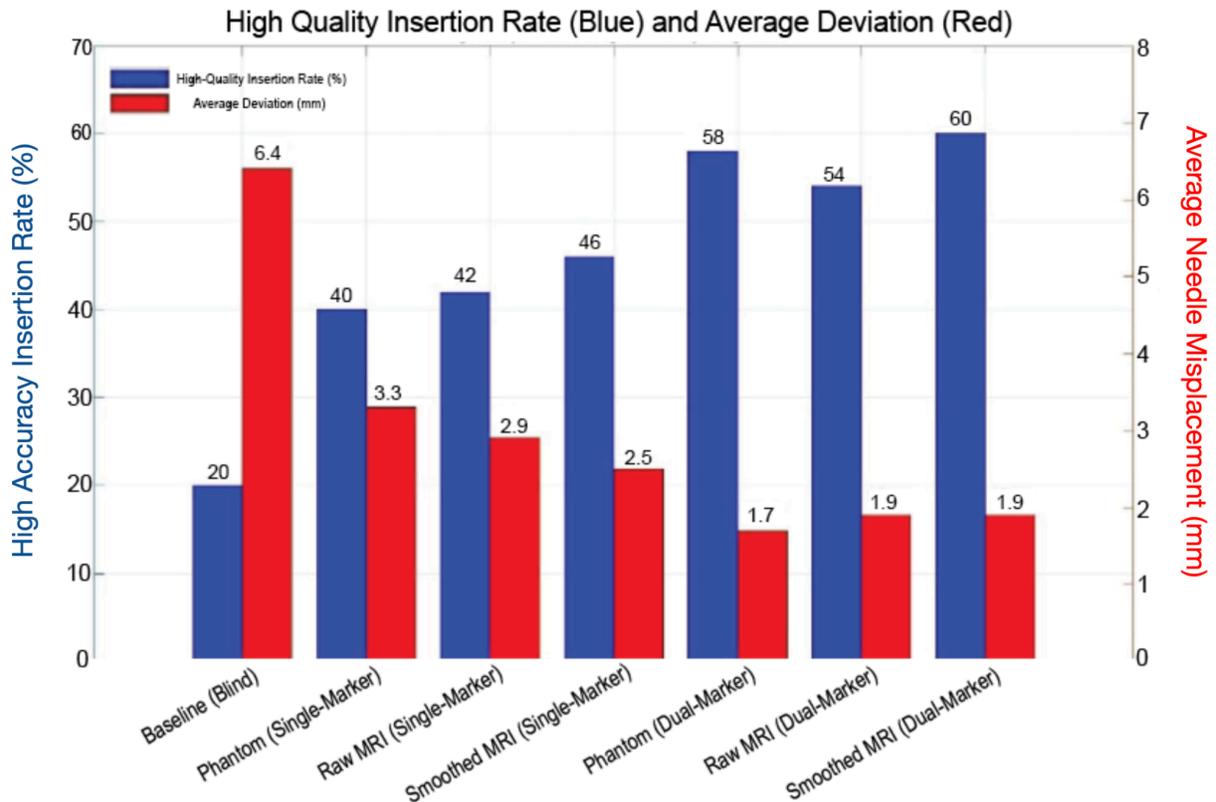

**Fig. 9:** High-accuracy insertion rate (blue bars on the left) and average needle misplacement (red bars on the right) for baseline (blind) and AR-guided trials (phantom, raw MRI, smoothed MRI).

The detailed calculations from these experiments are provided in the supplementary materials (Supplementary Tables 2 and 3).

## Discussion:

In this paper, we presented an MRI-based augmented reality (AR) surgical navigation system enhanced by dual-ArUco marker tracking. The system uses a Laplacian smoothing algorithm that finetunes the MRI spinal model. We observed that Laplacian smoothing increases the surface-shell Dice coefficient, generating a mesh that closely matches the phantom, thus permitting accurate AR overlays. Our system demonstrated all steps required for a successful implementation in a medical device, providing a radiation-free alternative to traditional fluoroscopy.

Our experimental data shows that improved needle insertion accuracy was achieved with dual-marker tracking compared to single-marker tracking. Specifically, the dual-ArUco marker tracking system yielded a "high-accuracy" needle insertion rate of approximately 60% with an average misplacement of 1.9 mm, as compared to the single-marker approach, which achieved a 46% "high-accuracy" insertion rate with a 2.5 mm average deviation. The dual-ArUco marker tracking error is comparable to the approximately 2 mm deviation typically observed with conventional fluoroscopy methods. As shown in Figure 9, the blue bars on the left show the "high-accuracy insertion rate". The plot shows that the single-ArUco marker system is better than the baseline (no marker), and that the accuracy of dual-ArUco markers is better than the accuracy of a single-ArUco marker system. The bars on the right (red columns) show the average needle misplacement. We observe a decrease in the placement error of the single-ArUco marker system compared to the baseline, and a further decrease from the single-ArUco marker system to the dual-ArUco marker system. This improvement is the result of improved tracking accuracy of the dual-marker configuration, allowing for more reliable marker-pose estimations and more stable anatomical overlays. Moreover, dual-ArUco marker systems provide inherent redundancy. If one marker is temporarily obstructed, the system automatically switches to single-ArUco marker tracking to maintain uninterrupted visualization. Additionally, the dual-ArUco marker system dynamically adjusts the scale of the 3D model overlay which improves the accuracy of guidance.

Future work will explore expanding our AR system to include more than two ArUco markers. We hypothesize that this will further enhance the system robustness and ensure continuous accuracy even under challenging operating room conditions. Increasing the number of ArUco markers provides additional backup points in the case of obstructed images and improves overall tracking stability. Furthermore, future studies must investigate improved optimization methods for the smoothing of the MRI model. For instance, advanced 3D Laplacian smoothing algorithms should be tailored specifically for patient-specific anatomical variations to improve the accuracy of AR overlays. Additionally, the tracking of surgical instruments relative to anatomical targets should be implemented in our system. To address this, we plan to develop and integrate trackable surgical instruments with 3D ArUco markers into our AR system for real-time visualization of the needle tip relative to the target region. With these enhancements, we plan to perform studies to compare our MRI navigation system with conventional fluoroscopy.

One final point must be discussed in evaluating the accuracy of our MRI based system. This point is related to the position of the ArUco markers on the patient and the change in the position of the ArUco markers, relative to the anatomy of a patient, in the time between when the MRI scan is taken and the time when the epidural is administered by the surgeon. In addition, potential changes in the positioning of the ArUco markers relative to the anatomy of a patient must be evaluated because the standard MRI procedures capture images of the patients in a supine position, while the epidurals are performed with the patients in a prone position. Any positional errors occurring due these positional changes must be evaluated and may require adjustments in the CAD model or in the MRI protocol. These issues will be studied in future investigations.

**Acknowledgements:**

This work was in part supported by the Research Corporation for Science Advancement under grant #SA-ABI-2023-039b. The authors would like to acknowledge Dr. Farshad Ahadian from the Center for Pain Medicine at the University of California, San Diego for his insightful discussions. Also, we would like to thank Darin Tsui from the School of Electrical and Computer Engineering, Georgia Institute of Technology, and Ananya Rajan at Diality for many helpful discussions throughout this project. This manuscript has used the language refinement option of ChatGPT (OpenAI).

**References**

[1]: Evans L, O'Donohoe T, Morokoff A, Drummond K (2023) The role of spinal surgery in the treatment of low back pain. Med J Aust 218:40-45. https://doi.org/10.5694/mja2.51788

[2]: Leggett LE, Soril LJ, Lorenzetti DL, Noseworthy T, Steadman R, Tiwana S, Clement F (2014) Radiofrequency ablation for chronic low back pain: a systematic review of randomized controlled trials. Pain Res Manag 19:e146-53. https://doi.org/10.1155/2014/834369

[3]: Palmer WE (2016) Spinal Injections for Pain Management. Radiology 281:669-688. https://doi.org/10.1148/radiol.2016152055

[4]: Kopacz, Dan J. MD; Allen, Hugh W. MD. Comparison of Needle Deviation During Regional Anesthetic Techniques in a Laboratory Model. Anesthesia & Analgesia 81(3):p 630-633, September 1995. https://pubmed.ncbi.nlm.nih.gov/7653834/

[5] Sun, Zhijuan; Inskip, Peter D.; Wang, Jixian; Kwon, Deukwoo; Zhao, Yongcheng; Zhang, Liangan; Wang, Qin; Fan, Saijun. Solid cancer incidence among Chinese medical diagnostic x-ray workers, 1950–1995: Estimation of radiation-related risks. International Journal of Cancer 138(12):p 2875–2883, June 2016. | DOI: 10.1002/ijc.30036. https://pubmed.ncbi.nlm.nih.gov/26860236/

[6]: D. Tsui, M. Jo, B. Nguyen, F. Ahadian and F. E. Talke, "Optical Surgical Navigation: A Promising Low-cost Alternative*," 2023 45th Annual International Conference of the IEEE Engineering in Medicine & Biology Society (EMBC), Sydney, Australia, 2023, pp. 1-4, doi: 10.1109/EMBC40787.2023.10340384. https://ieeexplore.ieee.org/document/10340384

[7]: Koszulinski A, Sandoval JS, Vendeuvre T, Zeghloul S, Laribi MA. (2022) Comanipulation Robotic Platform for Spine Surgery With Exteroceptive Visual Coupling: Development and Experimentation. Journal of Medical Devices 16(4):041002. https://doi.org/10.1115/1.4054550

[8]: Chen X, Tuncali K, Slocum AH, Walsh CJ. (2011) Design of an Instrument Guide for MRI-Guided Percutaneous Interventions. Journal of Medical Devices 5(2):027527. https://doi.org/10.1115/1.3590861

[9]: Pérez-Pachón L, Poyade M, Lowe T, Gröning F. Image Overlay Surgery Based on Augmented Reality: A Systematic Review. Adv Exp Med Biol. 2020;1260:175-195. doi: 10.1007/978-3-030-47483-6_10. PMID: 33211313. https://pubmed.ncbi.nlm.nih.gov/33211313/

[10]: Chen Y, Kwok KW, Ge J, Hu Y, Fok MP, Nilsson KR, Tse ZTH. (2014) Augmented Reality for Improving Catheterization in MRI-Guided Cardiac Electrophysiology Therapy. Journal of Medical Devices 8(2):020917. https://doi.org/10.1115/1.4027017

[11]: Li Y, Ma S, Yang Z, Jiang S, Lin Z, Zhou Z. (2024) A Multi-Optical and Mechanical Compensation Robotic Surgery System Based on Augmented Reality for Endoscopic Neurosurgery. Journal of Medical Devices 19(2):021005. https://doi.org/10.1115/1.4067172

[12]: Vávra, P., Roman, J., Zonča, P., Ihnát, P., Němec, M., Kumar, J., Habib, N., El-Gendi, A., Recent Development of Augmented Reality in Surgery: A Review, Journal of Healthcare Engineering, 2017, 4574172, 9 pages, 2017. https://doi.org/10.1155/2017/4574172

[13]: Hui J, Lu S, Lee E, Talke FE. (2024) Enhancing minimally invasive spinal procedures through computer vision and augmented reality techniques. J Biomed Eng Biosci 11:44–50. https://doi.org/10.11159/jbeb.2024.006

[14] Filipchuk, O. V., & Gurov, O. M. (2016). PECULIARITIES OF APPLYING BALLISTIC GEL AS A SIMULATOR OF HUMAN BIOLOGICAL TISSUES. Theory and Practice of Forensic Science and Criminalistics, 15, 367-373. https://doi.org/10.32353/khrife.2015.46

[15]: Fan F, Kreher B, Keil H, Maier A, Huang Y. Fiducial marker recovery and detection from severely truncated data in navigation-assisted spine surgery. Med Phys. 2022 May;49(5):2914-2930. doi: 10.1002/mp.15617. Epub 2022 Mar 31. PMID: 35305271. https://pubmed.ncbi.nlm.nih.gov/35305271/

[16]: Vela C, Fasano G, Opromolla R (2022) Pose determination of passively cooperative spacecraft in close proximity using a monocular camera and AruCo markers. Acta Astronautica 201:22-38. https://doi.org/10.1016/j.actaastro.2022.08.024

[17]: Vollmer, J. et al. "Improved Laplacian Smoothing of Noisy Surface Meshes." Computer Graphics Forum 18 (1999): n. Pag. https://www.ljll.fr/~frey/papers/meshing/Vollmer%20J.,%20Improved%20Laplacian%20smoothing%20of%20noisy%20surface%20meshes.pdf

[18]: International Journal of Innovations in Engineering and Technology (IJIET) Comparison of Jaccard, Dice, Cosine Similarity Coefficient To Find Best Fitness Value for Web Retrieved Documents Using Genetic Algorithm. https://dknmu.org/uploads/file/6842.pdf

[19] Solberg, Joseph; Copenhaver, David; Fishman, Scott M.. Medial branch nerve block and ablation as a novel approach to pain related to vertebral compression fracture. Current Opinion in Anaesthesiology 29(5):p 596-599, October 2016. | DOI: 10.1097/ACO.0000000000000375. https://pubmed.ncbi.nlm.nih.gov/27548307/

**Supplementary Table 1**

**Table S1.** Acquisition parameters for T1-weighted Dixon MRI sequence.

| Parameter | Value |
|---|---|
| Repetition time | Minimum allowed, usually $\geq 5.6$ ms |
| First echo time | 2.4 ms |
| Echo spacing | 1.2 ms |
| Number of echoes | 1 |
| Echo train length | 2 |
| Number of shots | 2 |
| Flip angle | $5^0$ |
| Slice thickness | 1 mm |
| Number of slices | 208 |
| Matrix (frequency direction) | 320 |
| Matrix (phase direction) | 162 |
| Field of view | 327mm |
| Pixel Bandwidth | 820 Hz |
| Acceleration | None |

**Table S2:** Injection Accuracy of Baseline and Single-ArUco Marker Experiments

|  | Blind (no marker) | Ground Truth Model (CAD) | MRI | Smoothed MRI |
|---|---|---|---|---|
| **Detector 1 (<1 mm)** | 10 | 20 | 21 | 23 |
| **Detector 2 (<5 mm)** | 18 | 20 | 19 | 20 |
| **Detector 3 (<10 mm)** | 12 | 8 | 10 | 7 |
| **Detector 4 (<20 mm)** | 10 | 2 | 0 | 0 |
| high-accuracy Insertion Rate (%) | 20% | 40% | 42% | 46% |
| Average Deviation (mm) | 6.4 | 3.3 | 2.9 | 2.5 |

**Table S3:** Injection Accuracy of Dual-ArUco Marker Experiments

|  | Ground Truth Model (CAD) | MRI | Smoothed MRI |
| --- | --- | --- | --- |
| **Detector 1 (<1 mm)** | 29 | 27 | 30 |
| **Detector 2 (<5 mm)** | 18 | 20 | 15 |
| **Detector 3 (<10 mm)** | 3 | 3 | 5 |
| **Detector 4 (<20 mm)** | 0 | 0 | 0 |
| high-accuracy Insertion Rate (%) | 58% | 54% | 60% |
| Average Deviation (mm) | 1.7 | 1.9 | 1.9 |